\documentclass{eurogen2023}
\usepackage{amsmath, amsfonts}
\usepackage{mathtools}
\usepackage{bm, bbm}
\usepackage{nicefrac}
\usepackage[noabbrev]{cleveref}
\usepackage{units}
\usepackage{subcaption}
\usepackage{float}

\newcommand{\x}{\bm{x}}
\newcommand{\z}{\bm{z}}

\title{GENERALIZATION CAPABILITIES OF CONDITIONAL GAN FOR TURBULENT FLOW UNDER CHANGES OF GEOMETRY}

\author{Claudia Drygala$^1$, Francesca di Mare$^2$, and Hanno Gottschalk$^1$}

\heading{Claudia Drygala, Francesca di Mare and Hanno Gottschalk}

\address{$^1$University of Wuppertal, School of Mathematics and Natural Sciences, IMACM \& IZMD \\
  Gaußstraße 20, D-42119 Wuppertal\\
  e-mail: \{drygala,hanno.gottschalk\}@uni-wuppertal.de\and
  $^2$
  Ruhr University Bochum, Department of Mechanical Engineering, Chair of Thermal Turbomachines and Aero Engines\\
  Universitätsstraße 150, D-44801 Bochum\\
  e-mail: francesca.dimare@ruhr-uni-bochum.de}


\abstract{Turbulent flow consists of structures with a wide range of spatial and temporal scales which are hard to resolve 
numerically. Classical numerical methods as the Large Eddy Simulation (LES) are able to capture fine details of turbulent 
structures but come at high computational cost. Applying generative adversarial networks (GAN) for the synthetic modeling 
of turbulence is a mathematically well-founded approach to overcome this issue. 

In this work, we investigate the generalization capabilites of GAN-based synthetic turbulence generators when geometrical changes 
occur in the flow configuration (e.g. \ aerodynamic geometric optimization of structures such as airfoils). As training data, 
we use the flow around a low-pressure turbine (LPT) stator with periodic wake impact obtained from highly resolved LES. 
To simulate the flow around a LPT stator, we use the conditional deep convolutional GAN framework pix2pixHD conditioned on 
the position of a rotating wake in front of the stator. For the generalization experiments we exclude images of wake positions 
located at certain regions from the training data and use the unseen data for testing. We show the abilities and limits of 
generalization for the conditional GAN by extending the regions of the extracted wake positions successively. Finally, we evaluate the statistical properties of the synthesized flow field by 
comparison with the corresponding LES results.}

\begin{document}

\section{INTRODUCTION}\label{sec:Introduction}
The structures of turbulent flows exhibit a wide range of spatial and temporal scales. Fine details of turbulent structures can be captured by classical numerical methods such as Large Eddy Simulation (LES), but they are associated with considerable computational costs. 
Recent developments in the field of Machine Learning (ML), make it possible to address this problem, where different approaches can be taken.

On the one hand, surrogate models, see e.g. \cite{Forrester2008}, attempt to learn maps from design parameters to objective functions. Other approaches propose to use machine learning for turbulence modeling, see e.g.\ \cite{Weatheritt2016}. More ambitious approaches use machine learning for the prediction of entire flow fields, cf.\ e.g.\ \cite{Singh2017}. King \emph{et al.} suggested to use generative advarsarial networks (GAN) to synthesize turbulent flows \cite{King2017}.

GAN synthesised solutions can be criticised as non physics based. Secondly, one could criticise that training data first needs to be produced at high computational cost to train the GAN, but one can use this data right away.  

In our previous work \cite{Drygala2022}, we addressed the first concern and proved that the application of GAN to ergodic systems is a mathematically well-founded approach, and that we are able to synthesize high quality turbulent flows with GAN, which well match the physics based features of LES. To answer to the second concern, we also made the first successful attempt to generalize with respect to spatial changes using the deep convolutional conditional GAN framework \texttt{pix2pixHD}. I.e.\ we generated tubulent flow configurations for geometries that have never been seen in the training of the GAN. 

The generalizability of GAN is one of the open questions for the synthesis of chaotic scenarios. Recently, this question has been addressed especially in the field of super-resolution reconstruction of turbulent (reactive) flows. In \cite{Bode2021}, the combination of fully and under-resolved data was used to improve the generalization capability, especially for extrapolation, for a physics-informed super-resolution GAN (SRGAN). The authors in \cite{Nista2022} investigated the generalization capabilities of their proposed SRGAN framework by changing parameters in the numerical setup such as the Reynolds number. In \cite{Deng2019}, the generalization capability of a SRGAN towards a more complicated wake flow configuration was investigated.

In this work, we investigate the generalization capabilities of synthetic turbulence generators with respect to geometric changes in the flow configuration. We use the flow around an low-pressure turbine blade (T106 LPT) with periodic wake impact obtained from high-resolution LES. We perform the synthesis of the turbulent flow using the conditional deep convolutional GAN framework \texttt{pix2pixHD} conditioned on the position of a rotating wake in front of the stator. For generalization experiments, we exclude wake positions located in certain regions from the training data and use the unseen data for testing. We investigate two strategies for constructing the data splits. We remove wake positions from the training data by excluding consecutive frames or by reducing the frame rate at regular intervals. We show the abilities and limitations of the 
generalization for the \texttt{pix2pixHD} by successively expanding the regions of the extracted wake positions and we compare the GAN synthesized and LES turbulent flows visually and by their statistical properties using two different physics-based evaluation metrics. Furthermore, we track the influence of the training data reduction on the computational cost.

\paragraph{Outline}
The paper is organized as follows. We briefly summarize the concept of ergodicity and describe the mathematical foundations behind conditional generative learning for ergodic systems in \cref{sec:Methodology}. This is followed by \cref{sec:Dataset}, where we introduce the training data and describe the construction of the data splits used for the generalization experiments. In \cref{sec:results}, we give details on the training of the GAN and the evaluation metrics and discuss the results of our numerical experiments. Finally, in \cref{sec:Conclusion_and_outlook} we give a conclusion and a short outlook. 

\section{METHODOLOGY}\label{sec:Methodology}

Our previous work \cite{Drygala2022} showed
that the application of generative learning for deterministic ergodic systems proves to be a mathematically well-founded approach
since it converges in the limit of large observation time. Below, we briefly recapitulate the notion of ergodicity for better understanding
and explain the mathematical foundations of conditional generative learning for ergodic systems. 

\subsection{Ergodicity}\label{ssec:Ergodicity}
The notion of image measure is crucial for the understanding of GAN and also for ergodicity. Let $\varphi : \Omega \to \Omega'$ be
a measurable mapping with respect to the $\sigma$-algebra $\mathcal{A}$ on $\Omega$ and $\mathcal{A}'$ on $\Omega'$. Then the image measure of a probalibity measure $\mu$ on $\mathcal{A}$ under a measurable mapping $\varphi:\Omega\to\Omega$ is defined by  
    \begin{align}
        \varphi_*\mu(A)=\mu(\varphi^{-1}(A)) ~~\forall A\in\mathcal{A}'.
    \end{align}
where $\varphi^{-1}(A)=\{x\in\omega|\varphi(x)\in A\}\in\mathcal{A}~\forall A\in\mathcal{A}'$.

We first define a probability space $(\Omega, \mathcal{A}, \mu)$ on a dynamic system as considered in this work. 
The state space $\Omega$ of a dynamic system
is given by a collection of mappings $\varphi^t:\Omega\to\Omega$ that satisfy $\varphi^0=\mathrm{id}_\Omega$ and 
$\varphi^t\circ\varphi^s=\varphi^{s+t}$ with $\varphi^t\circ\varphi^s(x)=\varphi^t(\varphi^s(x))~\forall x\in \Omega$.
In detail, the states $\varphi_t(x)$ at time $t\in\mathbb{R}$ can be derived by solving a (discretized) ordinary or 
partial differential equation starting in the initial state $x\in\Omega$. In our case of the numerical simulation 
of turbulent fluids, $\Omega=\mathbb{R}^d$ where $d$ is a large number of dimensions of the discretized state space 
of the fluid field. Moreover, the $\sigma$-Algebra $\mathcal{A}$ is a collection of events of the state space $A \subseteq \Omega$
and $\mu$ is a probability measure on $\mathcal{A}$. In this work, we investigate in particular the transformation of the
probability measure by a measurable mapping, also known as image measure. 
For the dynamical system, the probability measure $\mu$ is invariant if $\varphi^t_*\mu=\mu~\forall t\in \mathbb{R}$ holds, i.e.
all solution mappings $\varphi^t$ are measure preserving with respect to $\mu$. Without further mention, all mappings
are assumed to be measurable w.r.t. the suitable $\sigma$-algebra in the following. Lastly, we define the space 
$\mathcal{H}:=L^2(\Omega, \mathcal{A}, \mu)$ of all square-integrable functions $\psi:\Omega\to\mathbb{R}$ as the space of 
physical observables. 

Ergodicity of the dynamic system 
$\varphi_t$ w.r.t. an invariant measure $\mu$ is defined by 
\begin{align}
    \lim_{T\rightarrow \infty} \dfrac{1}{T} \int_{0}^{T} \psi\circ \varphi_t(x_0) \, \mathrm{d} t= \int_\Omega \psi(x) 
    \,\mathrm{d} \mu(x) =\mathbb{E}_ {\x \sim\mu}[\psi(\x)]~\forall x_0\in\Omega~.
    \label{eq:mean_ergodic_theorem_cont}
\end{align}
Hence, ergodicity equates the time average of a dynamic system with the ensemble average of its invariant measure \cite{Eisner2015, Birkhoff1931, Neumann1932}. 

In our numerical experiments, we do not train on data of the entire state-space $\Omega$ but we extract specific quantities related to trubulence.
This can be described by a mapping $\pi:\Omega\to\Omega'$ with $\Omega'$ the reduced state space. Even if the dynamics 
$\varphi_t$ can not be consistently formulated on the reduced state space $\Omega'$, ergodicity remains satisfied on $\Omega'$.
Let $\pi_*\mu$ be the projected measure. Assuming the ergodicty of the dynamic system on the entire state-space $\Omega$, 
we obtain
    \begin{align}
                \lim_{T\rightarrow \infty} \dfrac{1}{T} \int_{0}^{T} \psi\circ \pi\circ \varphi_t(x_0) \, 
                \mathrm{d} t= \int_{\Omega'} \psi(x') \, \mathrm{d} \pi_*\mu(x') ~\forall x_0\in\Omega
                \label{eq:mean_ergodic_theorem_cont_reduced}
    \end{align}
    when $f\circ \pi \in \mathcal{H}$. This easily follows from \eqref{eq:mean_ergodic_theorem_cont} and 
    $\int_{\Omega'} \psi(x') \, \mathrm{d} \,\pi_*\mu(x')=\int_{\Omega'} \psi\circ \pi(x) \, \mathrm{d} \,\mu(x)$, known as the general transformation formula. 

    Summarizing, GAN in turbulence doe not necessarily resolve the dynamics of the turbulent flow, but rather attempt to learn an representation of $\mu$ or $\pi_*\mu$, directly. As physical evaluations in the study of turbulence usually compute long time averages to capture the flow characteristics, the learned representation of $\mu$ or $\pi_*\mu$ enables the same evaluations and therefore a direct comparison to the numerical simulation.

\begin{figure}[t]
  \centering
    \includegraphics[width=\textwidth]{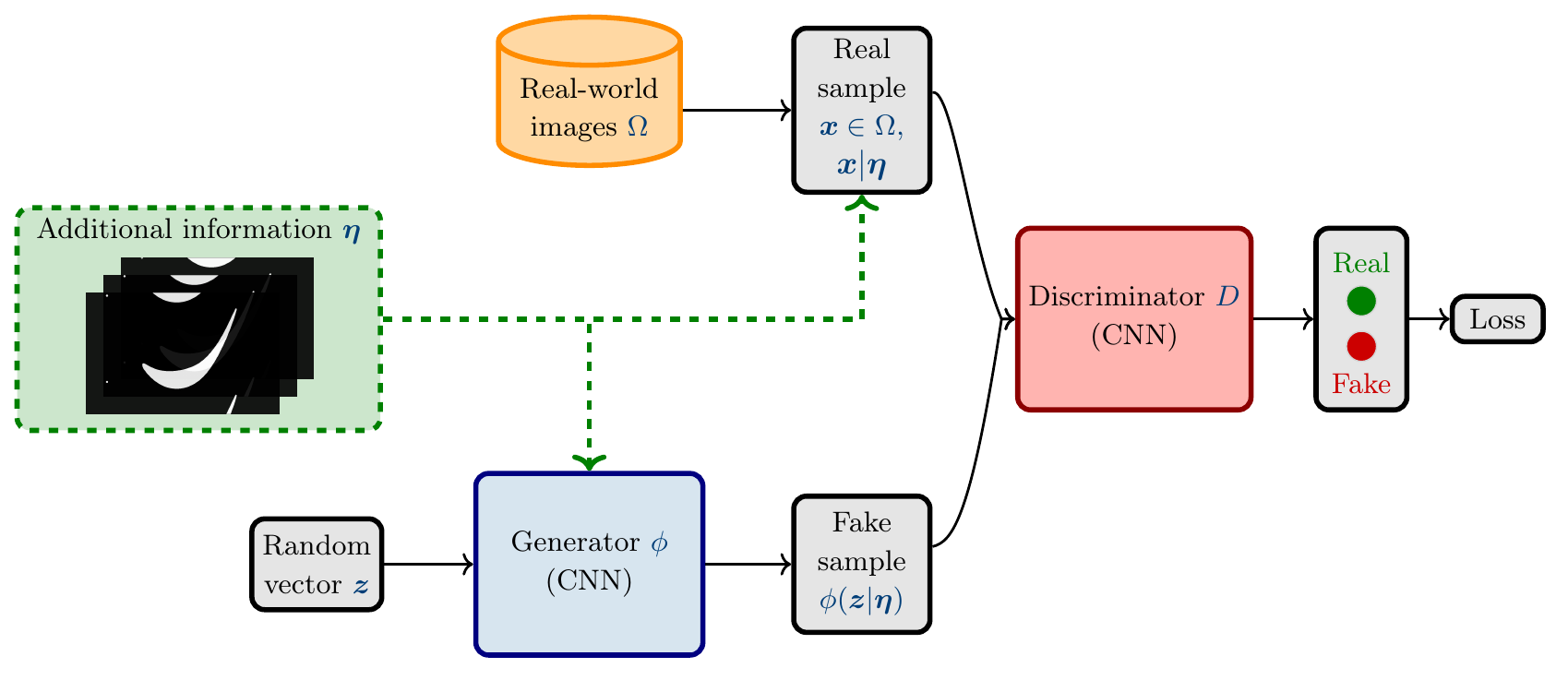}
    \caption{Architecture of a conditional GAN. According to: \cite{Goodfellow2020, cgan}. The generator produces fake samples 
    $\phi(\z|\eta) \sim \phi(\cdot|\eta)_*\lambda$ with the random noise vector $\z\sim \lambda$ conditioned on the additional information $\eta$ as its input. Here, $\eta$ is a binary segmentation mask
    that contains the position of a rotating wake in front of the stator. The discriminator receives both, fake 
    as well as real samples, and estimates the probability that the given input sample comes rather  
    from $\mu|\eta$ than from $\phi(\cdot |\eta)_*\lambda$. Hence, the output of $D$ is a single scalar value per sample 
    in the range of $[0,1]$. During training, the feedback from the discriminator reaches the generator when the weights of the GAN framework are updated by backpropagation. Here, the entire 
    GAN framework can be backpropagated in one go using the same loss function for $\phi$ and $D$. This is because both networks are fully differentiable and trained end-to-end. The problem \eqref{eq:cgan_loss} reaches its optimum when the generator captures the real-worlds distribution, through which the discriminator is unable to distinguish real from fake samples, i.e.  
    $\phi(\cdot |\eta)_*\lambda=\mu|\eta$  and $D(\cdot)=\nicefrac{1}{2}$.}
    \label{fig:cGAN}
\end{figure}

\subsection{Mathematical foundations of conditional generative learning for ergodic systems}\label{ssec:Conditional-generative-learning-for-ergodic-systems}

In this work, we synthesize turbulence using the conditional deep generative adversarial network (cDCGAN) 
~\texttt{pix2pixHD} \cite{cgan, dcgan, pix2pixHD}.
A cDCGAN consists basically of the two mappings $\phi :(\Lambda, \mathcal{V}) \to \Omega$ 
and $D: (\Omega, \mathcal{V}) \to [0,1]$ referred to as generator and discriminator. 
Here, $\Omega$ is a space containing the real-world data related to a family of unknown probability distribution $\{\mu|\eta\}_{\eta \in \mathcal{V}}$ and 
$\Lambda$ is a space of latent variables associated with a simple probability measure 
$\lambda$ like uniform or Gaussian noise. Furthermore, $\mathcal{V}$ is a space describing the additional information, on
which the GAN framework is conditioned, endowed with its corresponding probability distribution $\nu$. 
Thereby, the additional information can be provided, for example, by class labels or (semantic) segmentation masks
as in our case (see \cref{fig:cGAN}).
Conditioning a GAN framework with supplementary information makes it possible to control the data production process 
performed by the generator which transforms the noise measure $\lambda$ conditioned on $\eta$ which is a sample of the random variable $ \bm{\eta} \sim \nu$  to the image measure $\phi(\cdot |\eta)_*\lambda$.

The goal of adversarial learning is to learn a mapping $\phi$ from the feedback of the discriminator 
$D$ such that $D$ is unable to distinguish synthesized samples of $\phi_*(\lambda|\eta)$ from real-world data sampled 
from the conditioned target measure $\mu|\eta$. During training, $D$ learns to classify if an input sample comes rather 
from $\mu|\eta$ than from $\phi_*(\lambda|\eta)_*\lambda$ by assigning real-world data a high probability of being from $\mu|\eta$  
and synthesized data a low probability. 
Generative learning is successful, if $\phi$ is so well trained, that even the best discriminator is unable to distinguish 
between samples from $\mu|\eta$ and $\phi_*(\lambda|\eta)$. 

In case of the cDCGAN, the generator and discriminator are defined as convolutional neural networks (CNN) \cite{dcgan} which are well-known for their
successful application in the field of image processing \cite{cnn_impact, cnn_explanation}. 
In order to train the weights of $\phi$, the feedback
of $D$ for $\phi$ is performed by backpropagation \cite{backprop} through the concatenated mapping $D \circ \phi$.
Here, the universal approximation property of (deep) neural networks 
guarantees that any mappings $\phi$ and $D$ can be represented with a given precision under the condition that the 
architecture of the networks is sufficiently wide and deep \cite{asatryan2020convenient}.

The training is structured as two-player min-max game between the discriminator and the generator by solving the 
optimization problem  

\begin{align}
    \min_\phi \max_D \mathcal{L}_{\text{cond.}}(D, \phi)
    \label{eq:pix2pix_optimization_problem}
\end{align}
with the binary cross-entropy \cite{bce} as loss function 
\begin{align}
    \mathcal{L}_{\text{cond.}}(D, \phi) &= \mathbb{E}_{{\x\sim \mu\atop\bm{\eta}\sim \nu}}[\log(D(\x|\bm{\eta}))]
    +\mathbb{E}_{{\z\sim \lambda\atop \bm{\eta}\sim\nu}}[\log(1-D(\phi(\z|\bm{\eta})))]
    \label{eq:cgan_loss}
\end{align}
where $\mathbb{E}$ is denoting the expected value. 

In this work, we investigate a special type of cDCGAN called \texttt{pix2pixHD} \cite{pix2pixHD} which allows us to generate 
high-resolution photo-realistic images from semantic segmentation masks by modifying the architecture of $\phi$ and $D$
and extending the loss-function \eqref{eq:cgan_loss}.
The generator $\phi$ is composed of two subnetworks $\phi_1$ and $\phi_2$ assuming the role of a global generator 
and a local enhancer. This results in a coarse-to-fine generator $\phi=\{\phi_1, \phi_2\}$ aggregating the global and 
local information effectively. Instead of a single discriminator $D$, three multi-scale discriminators $D_1, D_2$ and $D_3$
are installed into the \texttt{pix2pixHD} framework, which have an identical network architecture
but operate with three different image resolutions. This leads to the extended optimization problem
\begin{align}
  \min_\phi \max_{D_1, D_2, D_3} \sum_{i=1}^3 \mathcal{L}_{\text{cond.}}(\phi, D_i)~.
  \label{eq:pix2pixHD_optimization_problem_multi_scales}
\end{align}
By downsampling the input images of the certain discriminators by factor two and four, the \texttt{pix2pixHD} creates 
a pyramid of images during the training. The discriminator operating on the coarsest scale has the largest receptive 
field which makes it is possible to guide the generator $\phi$ producing globally consistent images. 
By the discriminator working on the finest scale, 
the generator's attention can be directed to finer details in the data production.
Lastly, the introduction of a feature matching loss $\mathcal{L}_{FM}$ defined as
\begin{align}
    \mathcal{L}_{FM} = \mathbb{E}_{{\x\sim \mu\atop\bm{\eta}\sim \nu}}\left[\sum_{j=1}^L \frac{1}{N_j}
    \left( \lVert D_i^{(j)}(\x|\bm{\eta}) - D_i^{(j)}(\phi(\z|\bm{\eta}))  \rVert_1 \right)\right]
\end{align}
stabilizes the training of the \texttt{pix2pixHD}. Here, $D_i^{(j)}$ represents the $j$th-layer feature extractor \cite{feature_extraction} of the discriminator $D_i, ~i=1, 2, 3$, $L$ defines the number of layers, $N_j, ~ j=1, \ldots, L$ provides the number of elements contained in each layer and $\lVert \cdot \rVert_1$ denotes the $l_1$-norm.
Hence, the final optimization problem can be defined as
\begin{align}
  \min_\phi \left[ \left(\max_{D_1, D_2, D_3} \sum_{i=1}^3 \mathcal{L}_{\text{cond.}}(\phi, D_i) \right) + \gamma \sum_{i=1}^3 \mathcal{L}_{FM}(\phi, D_i) \right]
\end{align}
 where $\gamma$ is the weighting parameter.

\section{DATA GENERATION}
\label{sec:Dataset}

\subsection{Setup of experiments} \label{ssec:Setuo}
We perform our experiments on the test case of a flow around an academic low-pressure turbine blade (T106 LPT) with 
periodic wake impact obtained from highly resolved Large Eddy Simulations (LES) \cite{froehlichLES, surveyLES}. Here, the wakes are artificially generated by an upstream rotating bar grid and convected into the 
stator passages where the rotation of the flow within the passage leads to their deformation. In addition, the interaction 
between the periodically detaching boundary layer and the wakes occuring in the rear region of the LPT stator suction side
makes this test case to an interesting example of complex turbulent interactions.

\begin{figure}[ht]
\centering
        \begin{subfigure}[b]{0.47\textwidth}
	    \includegraphics[width=\textwidth]{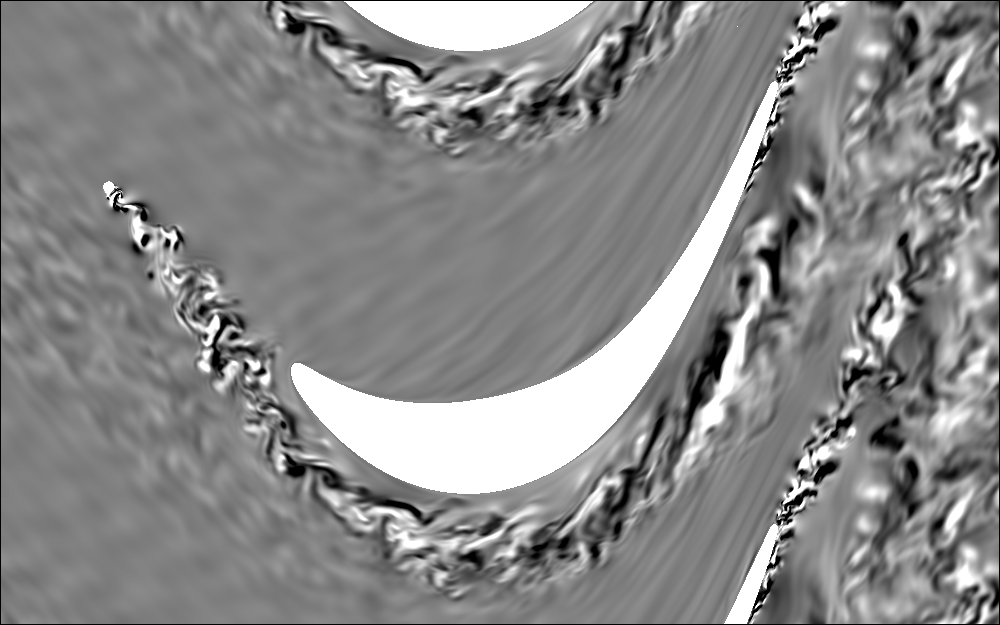}
	    \caption{T106 turbine stator}
        \label{fig:example-lpt-rgb}
    \end{subfigure}%
    \hspace{5mm}%
	\begin{subfigure}[b]{0.47\textwidth}
		\includegraphics[width=\textwidth]{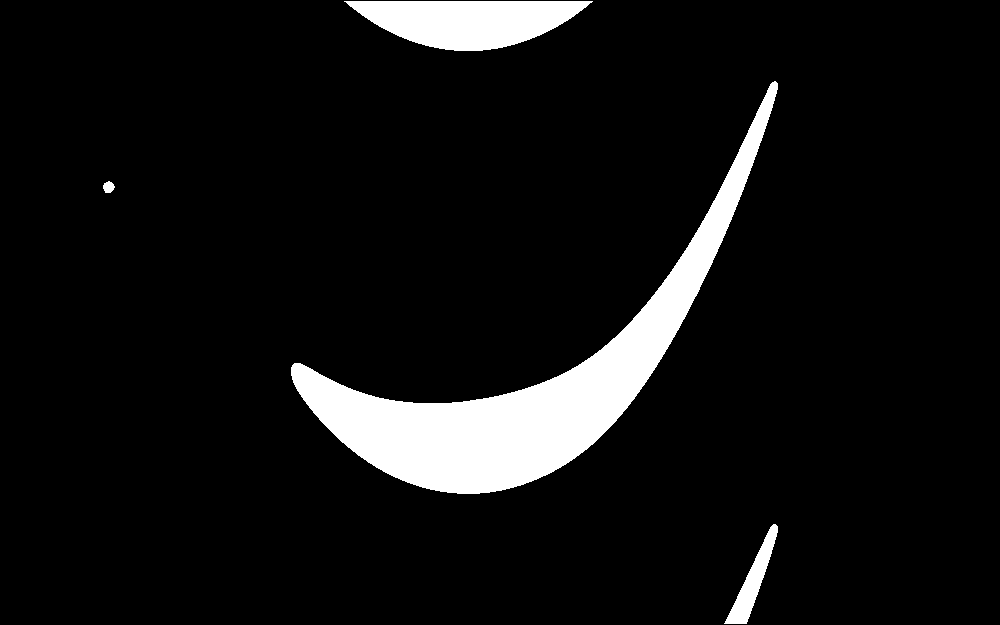}
		\caption{Binary segmentation mask}
		\label{fig:example-lpt-mask}
	\end{subfigure}
	\caption{A grayscale image for turbulence in the setting of a T106 turbine blade (\cref{fig:example-lpt-rgb} and its corresponding binary segmentation mask \cref{fig:example-lpt-mask}.}
    \label{fig:example-lpt}
\end{figure}

Post-processing the transient LES velocity field data generates the grayscale images (see \cref{fig:example-lpt-rgb}) used for GAN training
and statistical evaluation. Here, in the sense of \eqref{eq:mean_ergodic_theorem_cont_reduced}, a projection mapping is chosen
that represents the velocity component perpendicular to the image denoted by $w(\xi,t)$. 
In the upper left corner is the gray scale for $w(\xi,t)\approx0$. 
Negative values for $w(\xi,t)$ are displayed in lighter gray and positive values in darker grey. Hence, the smaller the value for $w(\xi,t)$ is, the lighter the gray becomes, and vice versa. 

The numerical setup of the LES can be found in \cite{Drygala2022}. In total, the data set consists of $2,250$ images, corresponding to 10 bar passing periods, with a resolution of $1,000 \times 625$ pixels. 

In addition to the grayscale images, the binary segmentation masks (see \cref{fig:example-lpt-mask}) are needed for both, GAN training and inference time which refers to the process of applying a trained generator to unseen data. These masks represent the position of the rotating wake in front of the stator.

\begin{figure}[ht]
\centering
        \begin{subfigure}[b]{0.325\textwidth}
	    \includegraphics[width=0.98\textwidth]{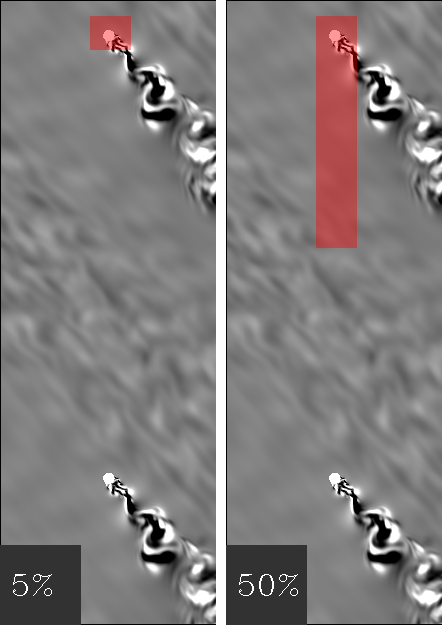}
	    \caption{Exclusion of sequential frames}
        \label{fig:example-exclusion-area}
    \end{subfigure}%
    \hspace{7mm}%
	\begin{subfigure}[b]{0.49\textwidth}
		\includegraphics[width=0.98\textwidth]{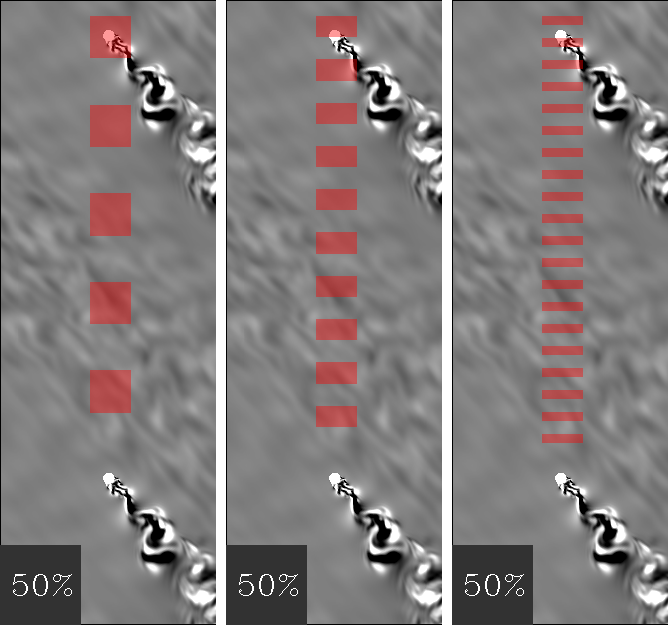}
		\caption{Reduction of the frame rate at regular intervals}
		\label{fig:example-reduction-framerate}
	\end{subfigure}
	\caption{Examples for the regions of the rotating wake positions excluded from the GAN training (marked by red boxes) in case a certain number of consecutive frames do not go into the GAN training (\cref{fig:example-exclusion-area}) or the frame rate is reduced at regular intervals (\cref{fig:example-reduction-framerate}).}
    \label{fig:example-generalization}
\end{figure}

\subsection{Computational cost}\label{ssec:Computational-cost-les}
The LES is performed on a partition of the High-Performance Computing (HPC) cluster of the Chair of Thermal Turbomachines and Aero Engines
featuring Intel Xeon ``Skylake'' gold 6132 CPUs with 2.6 GHz and 96 GB RAM.
For the simulation, $20$ nodes with 28 cores each had to be allocated and the computation time was approximately $8$ days, which corresponds to $640$ core weeks. 

\subsection{Data set splits}\label{ssec:Dataset-splits}
For the generalization experiments we exclude images of wake positions located at certain regions from
the training data and use the unseen data for testing. We study the abilities and limits of generalization for the cDCGAN by extending the regions of the extracted wake positions successively and hence reducing the variation in the training data.
The data set consists of $2,250$ images corresponding to $10$ bar passing periods as described above. We exclude a certain percentage from each of these periods from the training representing the test data. 
Here, we investigate two strategies for constructing the data splits. The first strategy is to extract consecutive frames starting at $5\%$ and increasing iteratively in steps of five percentage points up to $50\%$ (see \cref{fig:example-exclusion-area}. The second strategy is to extract a certain percentage of images starting at $20\%$, increasing iteratively in ten percentage point steps up to $90\%$, by reducing the frame rate at regular intervals, where the number of intervals varies from $5, 10$ and $20$. \Cref{fig:example-reduction-framerate} shows that the higher the number of intervals, the smaller the reduction of the frame rate at the certain region. This results in $10$ different data splits when excluding consecutive data frames, and $24$ different data splits when using the strategy of reducing frame rates at regular intervals. 
Additionally, we construct a test dataset consisting of 225 frames describing a full period for the evaluation of the physics-based metrics, in order to obtain a direct comparison and to consider how the lack of variation in the training data also affects the GAN synthesized turbulence of wake positions not excluded from the training.

\section{RESULTS OF EXPERIMENTS}\label{sec:results}
In this section we present and discuss the results of the numerical experiments. For this, we first provide the implementation details of the cDCGAN framework \texttt{pix2pixHD} setup and trained using PyTorch \cite{pytorch2019} . We compare the turbulence produced by GAN and LES on a visual level and consider their statistical properties, which are captured by physics-based metrics. Before going into the details of the results, we introduce these metrics in \cref{ssec:Physics-based-evaluation-metrics}. 

\subsection{Implementation details for cDCGAN training and inference}\label{ssec:Implementation-details-for-GAN}
For the experiments with the \texttt{pix2pixHD} framework, we use the original implementation \cite{pix2pixHD} with small modifications. We replace the reflection padding with a replication padding and add a replication padding to the global generator before the convolution during the downsampling procedure to prevent artifacts from appearing in the data synthesized by the generator $\phi$.
In the following, we often use the term inference to refer to the process of applying a trained generator to unseen data. 
As described in \cref{sec:Methodology} additional information $\bm\eta$ is incorporated into the cDCGAN training and at inference time. 
Here, $\bm\eta$ are binary segmentation masks (see \cref{sec:Dataset}), corresponding to a uniform distribution $\bm{\eta}\sim\nu_{\text{unif.}}$ over the wake coordinate $y$.
We train the \texttt{pix2pixHD} using the data splits described in \cref{ssec:Dataset-splits}. Training requires the grayscale images obtained from LES and the binary segmentation masks, while inference requires only the binary segmentation masks excluded from training. In practice, noise is not included in the \texttt{pix2pixHD} framework, since previous work shows that it is ignored while training \cite{pix2pix}. With regard to the image size $w \times h,~ w \neq h$, it must be taken into account for training purposes that $w$ and $h$ are divisible by $32$. For this reason, we resize the images for training to the size $w \times h = 992 \times 624$ while preserving the aspect ratio.
We train the \texttt{pix2pixHD} for all data set splits for $200$ epochs with a batch size of $10$ and a learning rate of $2 \times 10^{-4}$, updating the weights using the Adam optimizer \cite{ruder} with the parameters $\beta_1=0.9$ and $\beta_2=0.999$.

\subsection{Physics-based evaluation metrics}\label{ssec:Physics-based-evaluation-metrics}
We compare GAN synthesized and LES turbulence by quantities that can be cast in an abstract form $\mathbb{E}_ {\x \sim\mu}[\psi(\x)]$ of \eqref{eq:mean_ergodic_theorem_cont} and \eqref{eq:mean_ergodic_theorem_cont_reduced} given a certain evaluation function $\psi$.
Our previous work \cite{Drygala2022} shows that this approach is reasonable since any statistic evaluated on GAN synthesized flow fields converges on average to the corresponding statistic evaluated on LES data in the limit of large data and large network capacity, which makes this convergence uniform over all uniformly bounded functions $\psi$. 

\begin{figure}[t]
    \centering
    \includegraphics[width=0.8\textwidth]{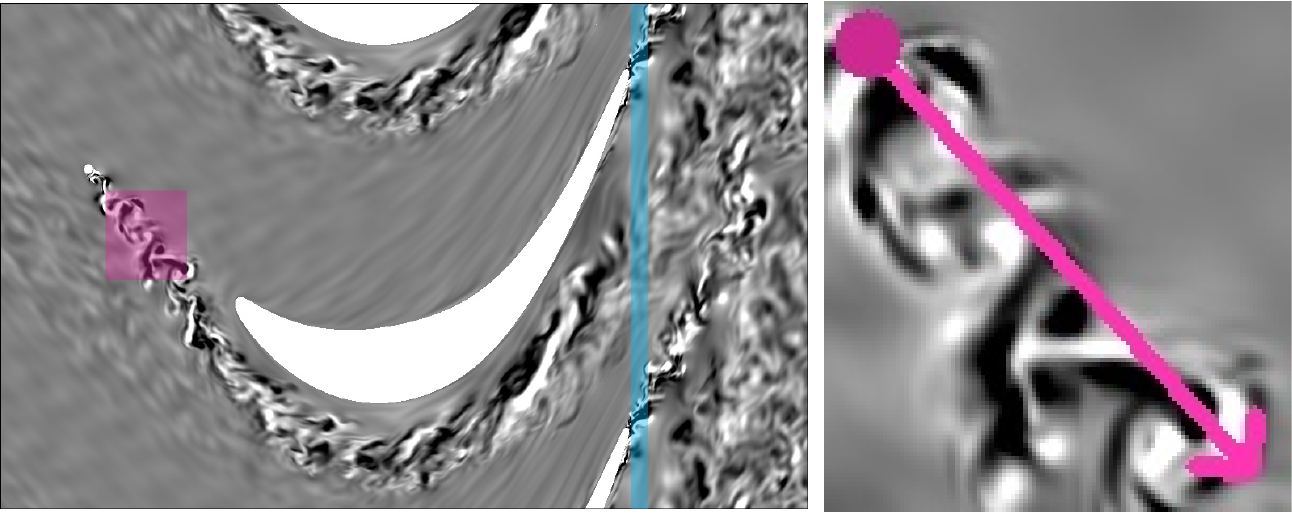}
    \caption{Areas examined to compare the moving average of the velocity field $z$-component for $w(\xi,t)>0$ (blue shaded) and the turbulent flow field correlations (pink shaded). In the close-up (right), the arrow shows the directory of the vector $p$ and the dot describes the co-moving pixel $\xi(\tau)$. }
    \label{fig:metric-computation-description}
\end{figure}

We evaluate the LES and GAN synthesized image $\x_\xi$ at pixel $\xi$ showing a snapshot of the $z$-component $V_z(\xi,t)$ of the velocity field at fixed time $t$ using two different metrics. For the turbulence patterns $\x_\xi=V_z(\xi)$ in front of the rotor blade we compute the correlation 
\begin{align}
\label{eq:correlation-coefficient}
    \rho_{V_z, V_z}(p) = \dfrac{\textup{cov}_\mu\left[V_z(\xi(\tau)), V_z(\xi(\tau)+p)\right]}{\sigma_{V_z(\xi(\tau))} \sigma_{V_z(\xi(\tau)+p)}}~,
\end{align}
where $\xi(\tau)$ is a point that is co-moving with the wake and $p$ is a vector pointing in the opposite direction of the vector connecting $\xi(\tau)$ to the wake.
Here, the co-moving pixel $\xi(\tau)$ is chosen to have a high degree of variation in $V_z(\xi(\tau))$ (see \cref{fig:metric-computation-description}). Furthermore, $\textup{cov}$ denotes the covariance defined by 
\begin{align}
\label{eq:covariance-vs-psi}
    \begin{split}
      &\textup{cov}_\mu\left[V_z(\xi(\tau)), V_z(\xi(\tau)+p)\right]\\
      &=\mathbb{E}_{\x \sim \mu_\tau}[\psi_{2,\xi(\tau),p}(\x)]-\mathbb{E}_{\x \sim \mu_\tau}[\psi_{1,\xi(\tau)}(\x)]\mathbb{E}_{\x \sim \mu_\tau}[\psi_{1,\xi(\tau)+p}(\x)]~,
    \end{split}
\end{align}
where $\mu_\tau$ is the limiting measure of the turbulent flow conditioned on the wake position at time $\tau$, using the evaluation functions $\psi_{1,\xi}(\x)=\x_{\xi}$ and $\psi_{2,\xi,p}(\x)=\x_{\xi}\x_{\xi+p}$. The standard deviation represented by $\sigma$ can be derived analogously using evaluation functions.

For the turbulence patterns $\x_\xi=V_z(\xi, t)$ in the rear region of the LPT stator suction side, we compute the moving average of the velocity magnitude for $w(\xi,t) > 0$. 
As described in \ref{sec:Dataset} the velocity component $w(\xi,t)>0$ is displayed by darker grey values and the gray scale for $w(\xi,t)\approx 0$ is in the upper left corner of the image.
To obtain the information about which pixel values of the image $\x_\xi$ satisfy the condition $w(\xi, t) > 0$, we apply an image thresholding technique \cite{image-thresholding}, which results in a binary image. Hence, we segment $\x_\xi$ using the rule 
\begin{align}
        s(\xi, t) = \begin{cases}
                ~~~0, ~V_z(\xi, t)> \epsilon(\xi, t)\\
                255, ~V_z(\xi, t) \leq \epsilon(\xi, t)
        \end{cases}
\end{align}
with $s(\xi, t)$ the pixel value of the binary image and $\epsilon(\xi, t)$ an adaptive threshold calculated at each pixel $\xi$ individually by computing the cross-correlation with a Gaussian window \cite{cross-correlation} over the neighbourhood of $\xi$ minus a constant $C$. The neighbourhood is defined by a quadratic structuring element \cite{Drygala-BGFG} of size $15 \times 15$ pixels with $\xi$ in the center and $C=V_Z(0,t)$ is the gray scale of the upper left corner.
Using the normalized values of $s(\xi, t)$ we evaluate the moving average $\psi(x)=x_\xi$ at pixel $\xi$ 
expressed by $\frac{1}{T}\int_0^T s(\xi,t)\,\textup{d} t=\mathbb{E}_{\x\sim\mu}[x_\xi]$ whereby this function is bounded on the normalized data.
The evaluations are performed over a small grid of 20 pixels in $x$-direction immediately after the rotor blade along the $y$-axis as shown in \cref{fig:metric-computation-description}.

\subsection{Results of excluding sequential frames}\label{ssec:results-seq-frames}
\begin{figure}[H]
  \centering
    \includegraphics[width=0.9\textwidth]{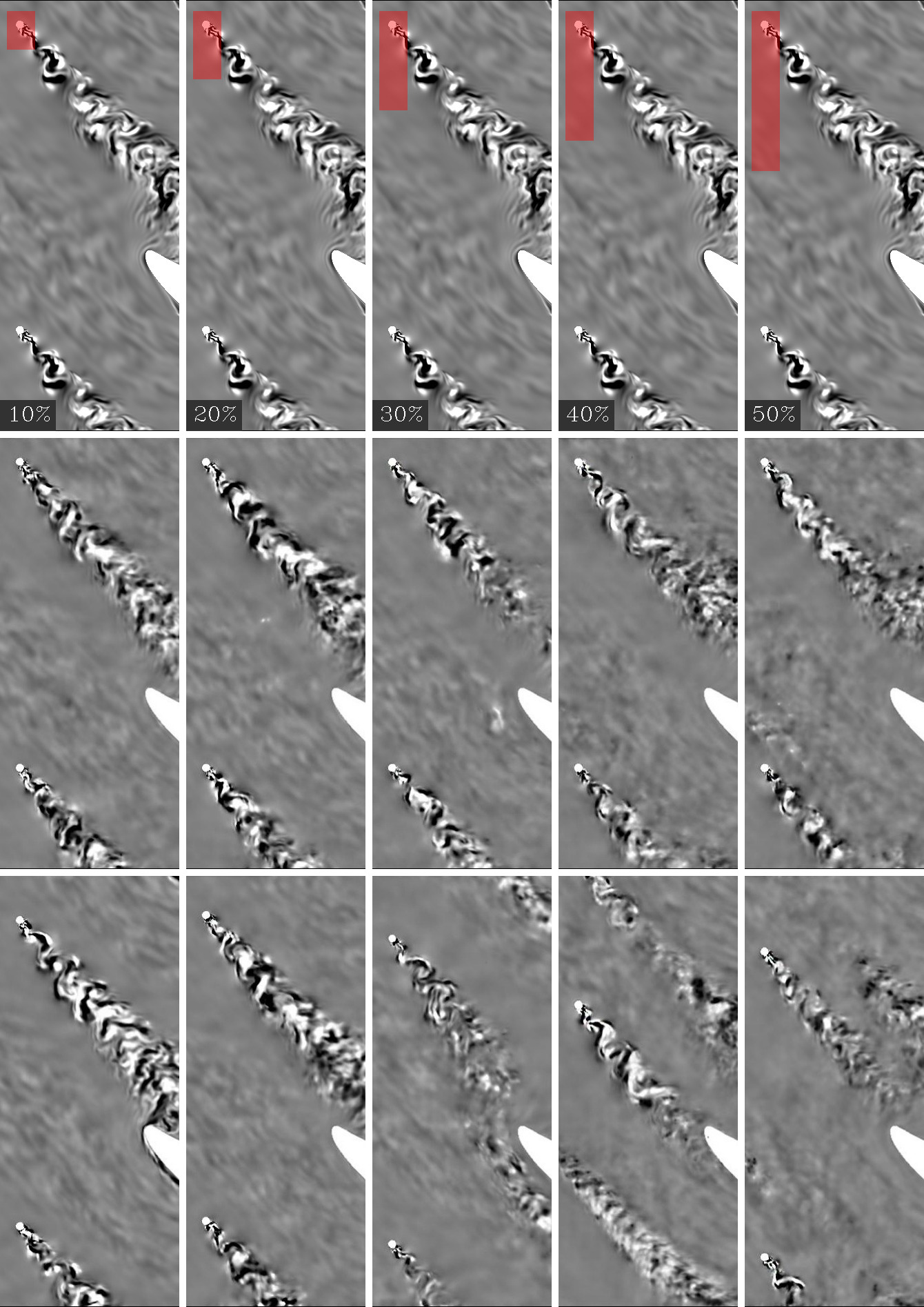}
    \caption{Comparison of LES (top row) and GAN synthesized (middle and bottom rows) turbulent flow fields, where the generator $\phi$ is not trained on the consecutive wake positions located in the red shaded region. The middle row is a direct comparison to the LES turbulence above and shows turbulence synthesized from wake positions located at the outer end of the extracted region. The bottom row shows GAN synthesized turbulence with the wake position in the center of  the extracted region.}
    \label{fig:results-consecutive-front}
\end{figure}

\begin{figure}[H]
    \centering
    \includegraphics[width=\textwidth]{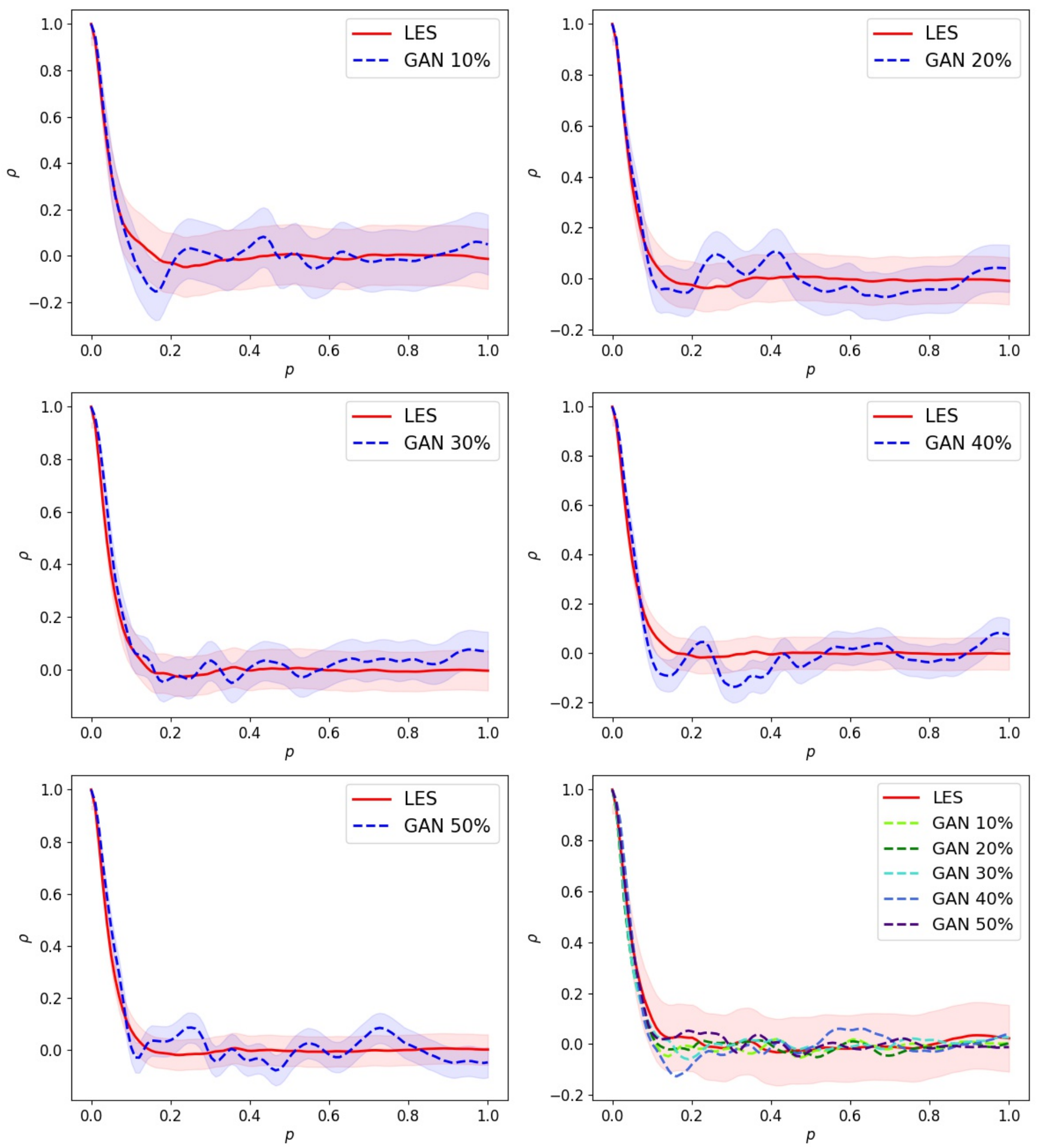}
    \caption{Comparison of the pointwise correlation for the LES and GAN synthesized turbulent flow fields, where the generator $\phi$ is not trained on a certain percentage of consecutive wake positions. The figure at the bottom row on the right shows the evaluation results for an entire period. The red and blue shaded areas indicate the $95\%$ confidence intervals of the respective curve.}
    \label{fig:correlation-res-consecutive}
\end{figure}

We start by examining the generated turbulence in front of the LPT stator. 
\Cref{fig:results-consecutive-front} shows that there are problems in the generation of turbulence starting from the exclusion of a whole region of 30\%. 
When excluding successive frames, it is significant that the more the wake position is located within the unseen training region, the more problems the GAN encounters in generating.  From the exclusion of 30\% of the training data, we can observe that the generated turbulence near the edge of the excluded area is of much better quality than those with wake positions in the center of the excluded region. Here, we see that turbulent flows disappear and that even artifacts are generated. These visual impressions are reflected in the correlations (see \cref{fig:correlation-res-consecutive}). 

The comparison of LES and GAN synthesized flow for the wake positions not seen in training shows, that the curves of the GAN synthesized images deviate from those of the LES curve especially by an exclusion of 40\% of the training data. Note that the test data sets become larger with the reduction of the training data, so that in the cases of 10\% and 20\% exclusion outliers can have a stronger effect on the evaluation metric. Therefore, we also compare the LES and GAN synthesized images over an entire period of $225$ images. 
Here, we see that especially the curve for the GAN synthesized images with 40\% less training data deviates from the LES curve. However, we also observe that all curves of the GAN synthesized images are within the confidence interval of the LES curve, indicating that the lower variation in training does not have a negative impact on the generation of turbulence for the wake positions included in the training.

\begin{figure}[t]
    \centering
    \includegraphics[width=0.8\textwidth]{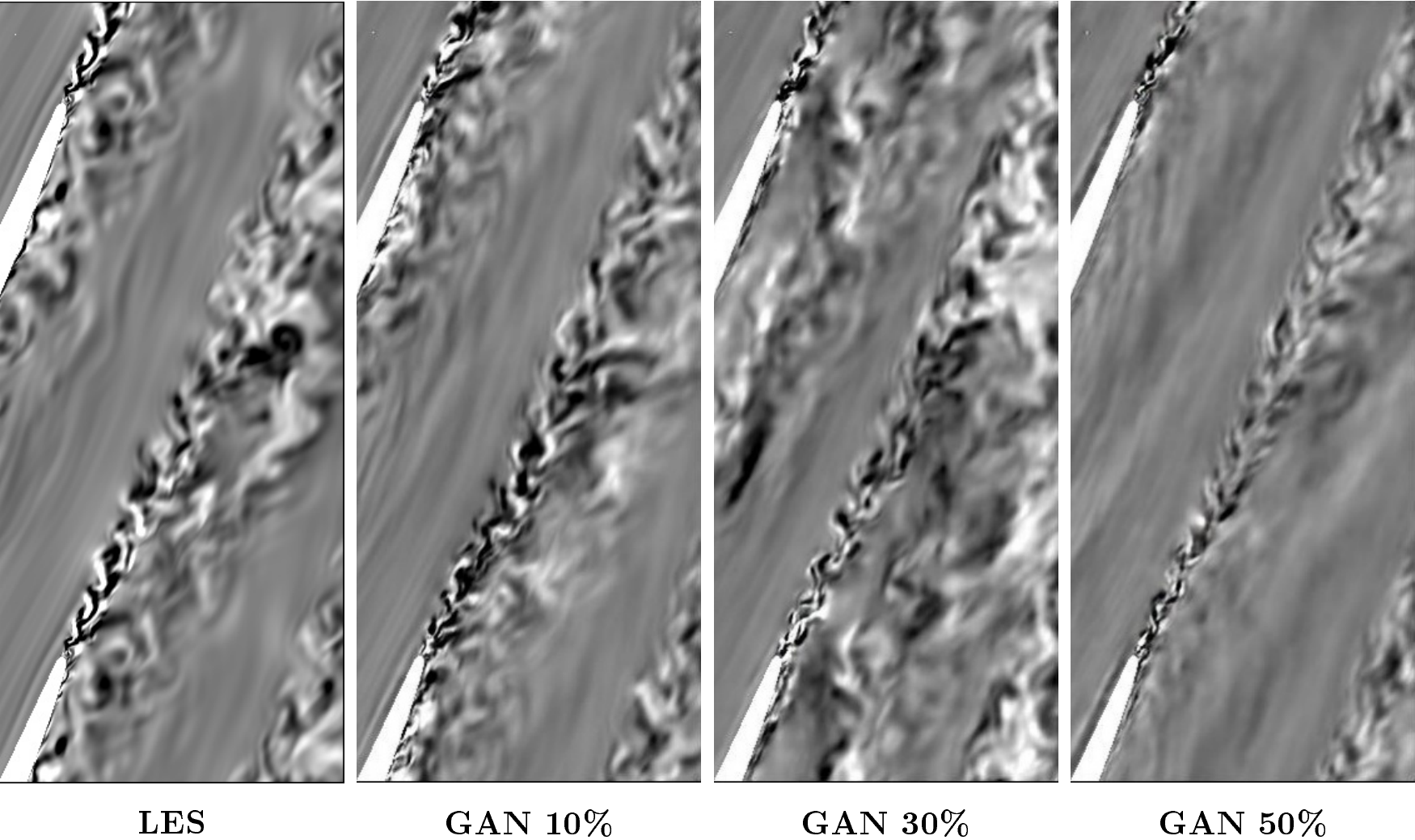}
    \caption{Comparison of LES and GAN synthesized turbulent flow in the rear region of the LPT stator suction side, where the generator $\phi$ is not trained on a certain percentage of consecutive wake positions.}
    \label{fig:results-consecutive-back}
\end{figure}

\begin{figure}[ht]
    \centering
    \includegraphics[width=\textwidth]{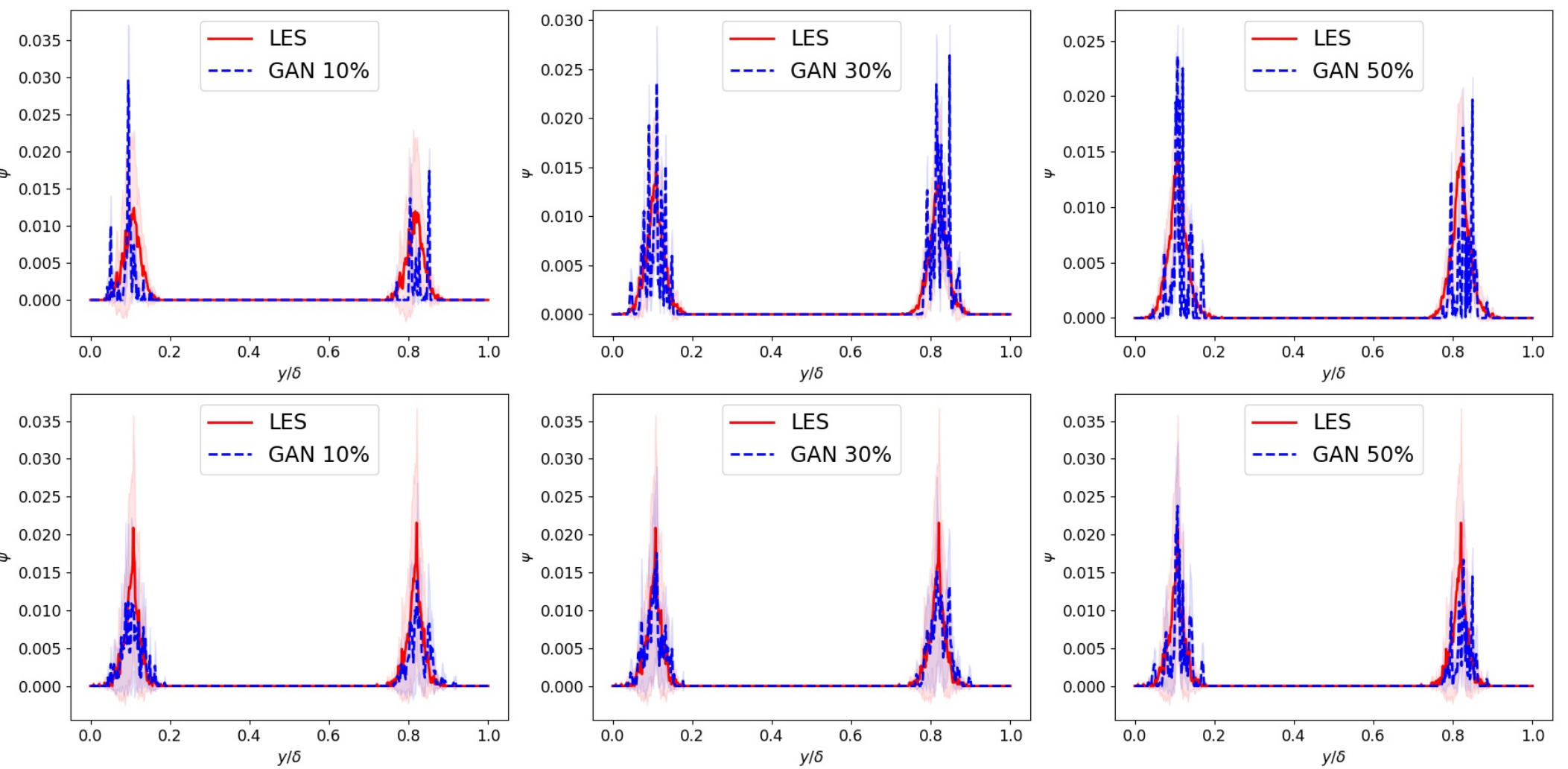}
    \caption{Comparison of the moving average of the velocity field $z$-component for $w(\xi,t)>0$ for LES and GAN synthesized turbulent flow in the rear region of the LPT stator suction side, where the generator $\phi$ is not trained on a certain percentage of consecutive wake positions. The top row shows the evaluation results for the wake positions which are excluded from the training, while the bottom row shows the results for an entire period.  The red and blue shaded areas indicate the $95\%$ confidence intervals of the respective curve.}
    \label{fig:ma-res-consecutive}
\end{figure}

A behavior similar to the turbulent flow in front of the LPT stator is observed for the GAN synthesized flow in the rear region of the LPT stator suction side. \Cref{fig:results-consecutive-back} shows that the quality of the GAN turbulence is excellent compared to the LES turbulence when the training data is reduced by $10\%$. The quality decreases reaching a reduction of $30\%$ for the training data by $\phi$ which generates artifacts. This decrease continues to grow and at the level of $50\%$ reduction of the training data, the generated turbulent flow is no longer incisive. 
The examination of the moving average $\psi$ of the velocity field $z$-component $w(\xi,t)>0$ (see \cref{fig:ma-res-consecutive}) reflects the visual impression. Except for a few outliers, the curve of the GAN synthesized turbulent flow lies within the $95\%$ confidence interval of the LES curve when the training data is reduced by $10\%$. The more the training data is reduced, the more confused the curve becomes. For a better comparison, we also evaluate $\psi$ on the GAN synthesized and LES turbulence of a full period. 
Here, we observe that the curve of the GAN synthesized turbulence is in excellent agreement with the curve of the LES, especially up to a reduction of $30\%$ of the training data. 
Thus, the reduction of the training variance has no negative impact on the turbulent flow in the rear region of the LPT stator suction side.  

\subsection{Results of frame rate reduction at regular intervals}\label{ssec:results-reduced-frame-rate}
The second setup for the generalization experiments is to reduce the training data by a certain percentage by reducing the frame rate at regular intervals, where the number of intervals varies from 5, 10, and 20. As described in \cref{ssec:Dataset-splits}, the consequence is that the higher the number of intervals, the smaller the reduction of the frame rate in a given region.

\begin{figure}[H]
  \centering
    \includegraphics[width=0.78\textwidth]{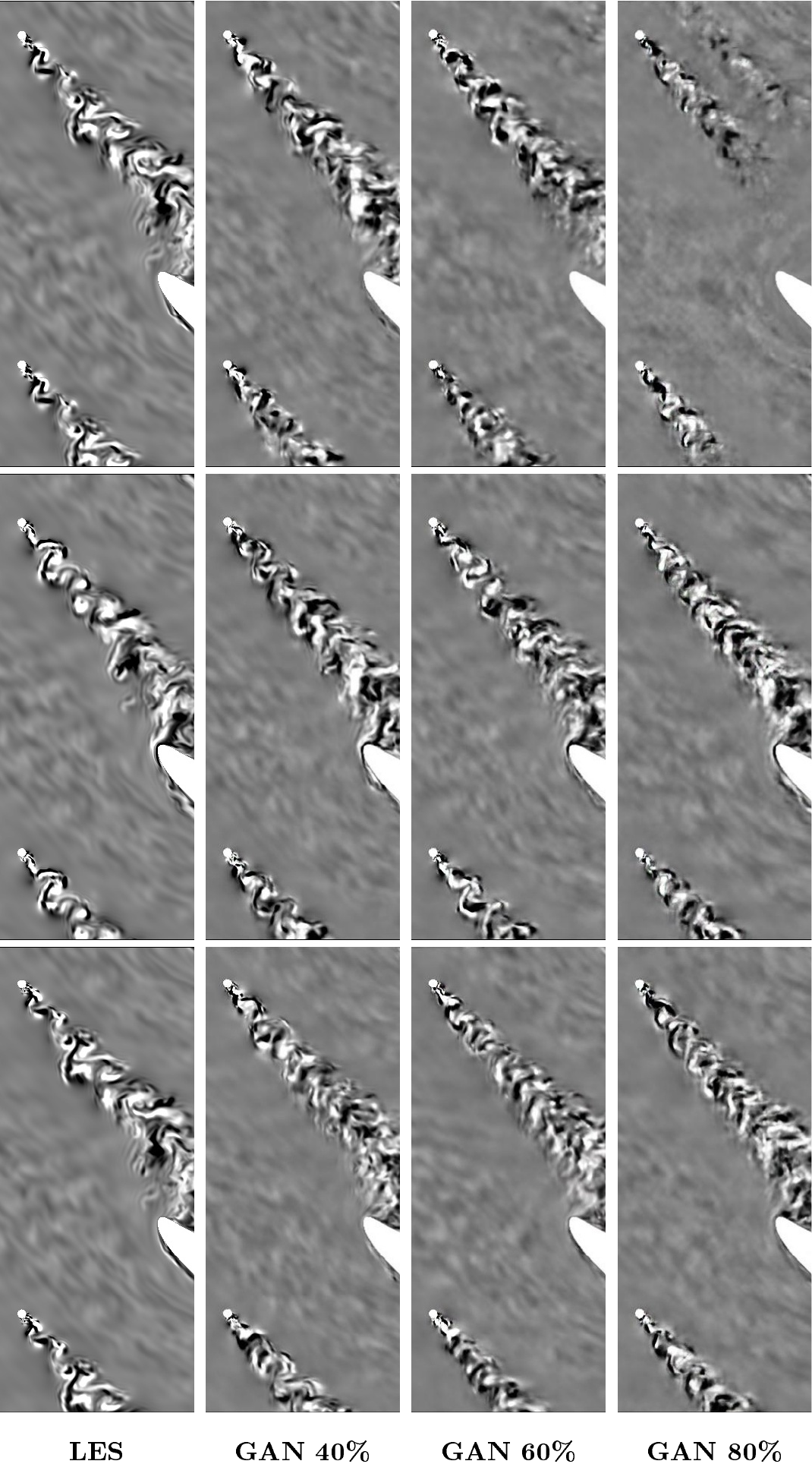}
    \caption{Comparison of LES and GAN synthesized turbulence, where the generator $\phi$ is trained on data sets with reduced frame rate at 5 (top), 10 (middle) and 20 (bottom) regular intervals, given a certain percentage of data to be excluded.}
    \label{fig:results-reduction-framerate}
\end{figure}

\begin{figure}[ht]
  \centering
    \includegraphics[width=\textwidth]{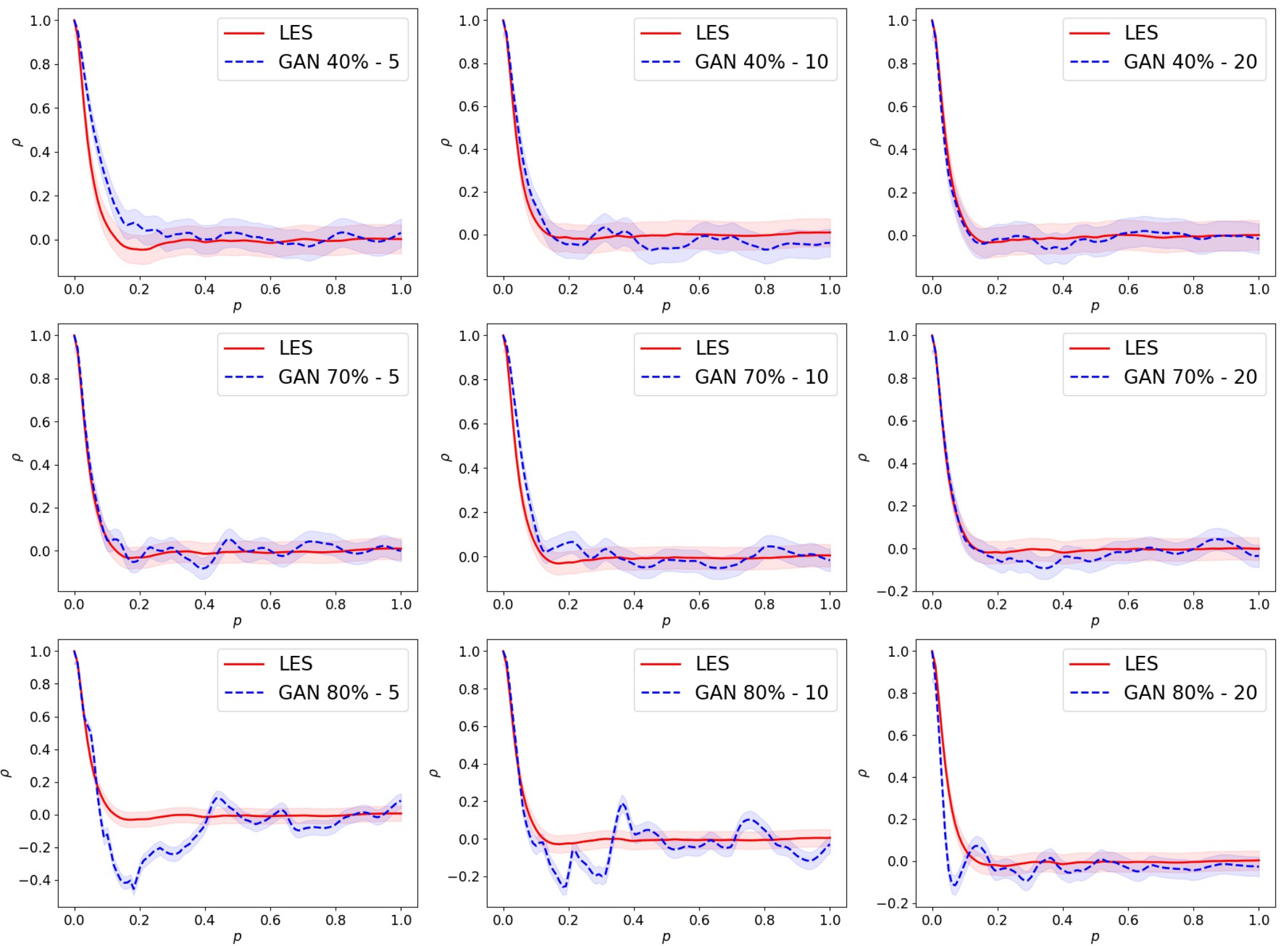}
    \caption{Comparison of the pointwise correlation for the LES and GAN synthesized turbulent flow fields, where the generator $\phi$ is trained on data sets with reduced frame rate at 5, 10 and 20 regular intervals, given a certain percentage of data to be excluded. The red and blue shaded areas indicate the $95\%$ confidence intervals of the respective curve.}
    \label{fig:correlation-reduction-framerate}
\end{figure}

\Cref{fig:results-reduction-framerate} shows that the choice of the number of intervals is of great importance. We see that the quality of the turbulent flows synthesized by GAN becomes much better as the number of intervals increases. For an interval number of 5, we obtain GAN synthesized images of poor quality, especially when the training data is reduced by 60\% and 80\%, since parts of the turbulent flow vanish and artifacts are generated. However, for the interval numbers of 10 and 20, we obtain GAN synthesized turbulence that can visually match that of LES when the training data is reduced by less than 80\%.
Taking a closer look, we can see that the turbulence patterns synthesized by the GAN trained on only 20\% of the original training data do not contain as much variation compared to the one of LES.
The correlation comparison of the LES and GAN synthesized turbulence mirrors the results at the visual level (see \cref{fig:correlation-reduction-framerate}). Up to a reduction of 40\% of the training data, the correlations of the GAN synthesized turbulence are in excellent agreement with those of the LES for an interval number of 20. Up to 70\%, only a slight decrease in performance can be observed. By reducing the training data to 80\%, we obtain completely different curves for the correlation of the GAN synthesized turbulence compared to the one of LES.
Here, the evaluation over an entire period is not shown additionally, since the proportions of the wake positions not seen in the training account for the largest part of the period at these high percentage values and the curves shown here are therefore already representative.

\begin{figure}[ht]
  \centering
    \includegraphics[width=0.8\textwidth]{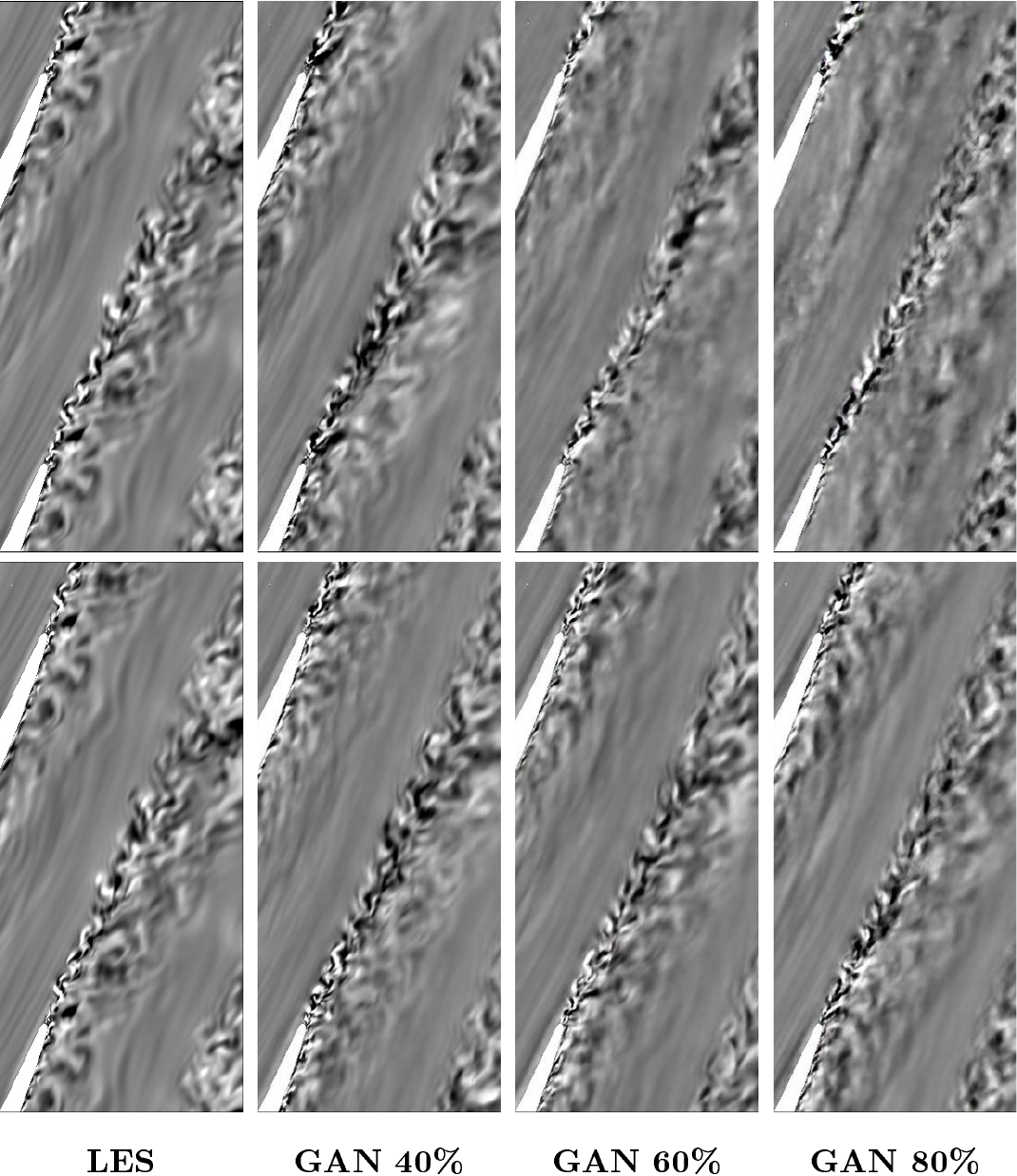}
    \caption{Comparison of LES and GAN synthesized turbulence in the rear region of the LPT stator suction side, where the generator $\phi$ is trained on data sets with reduced frame rate at 5 (top) and 20 (bottom) regular intervals, given a certain percentage of data to be excluded.}
    \label{fig:results-reduction-framerate-back}
\end{figure}

\begin{figure}[ht]
  \centering
    \includegraphics[width=\textwidth]{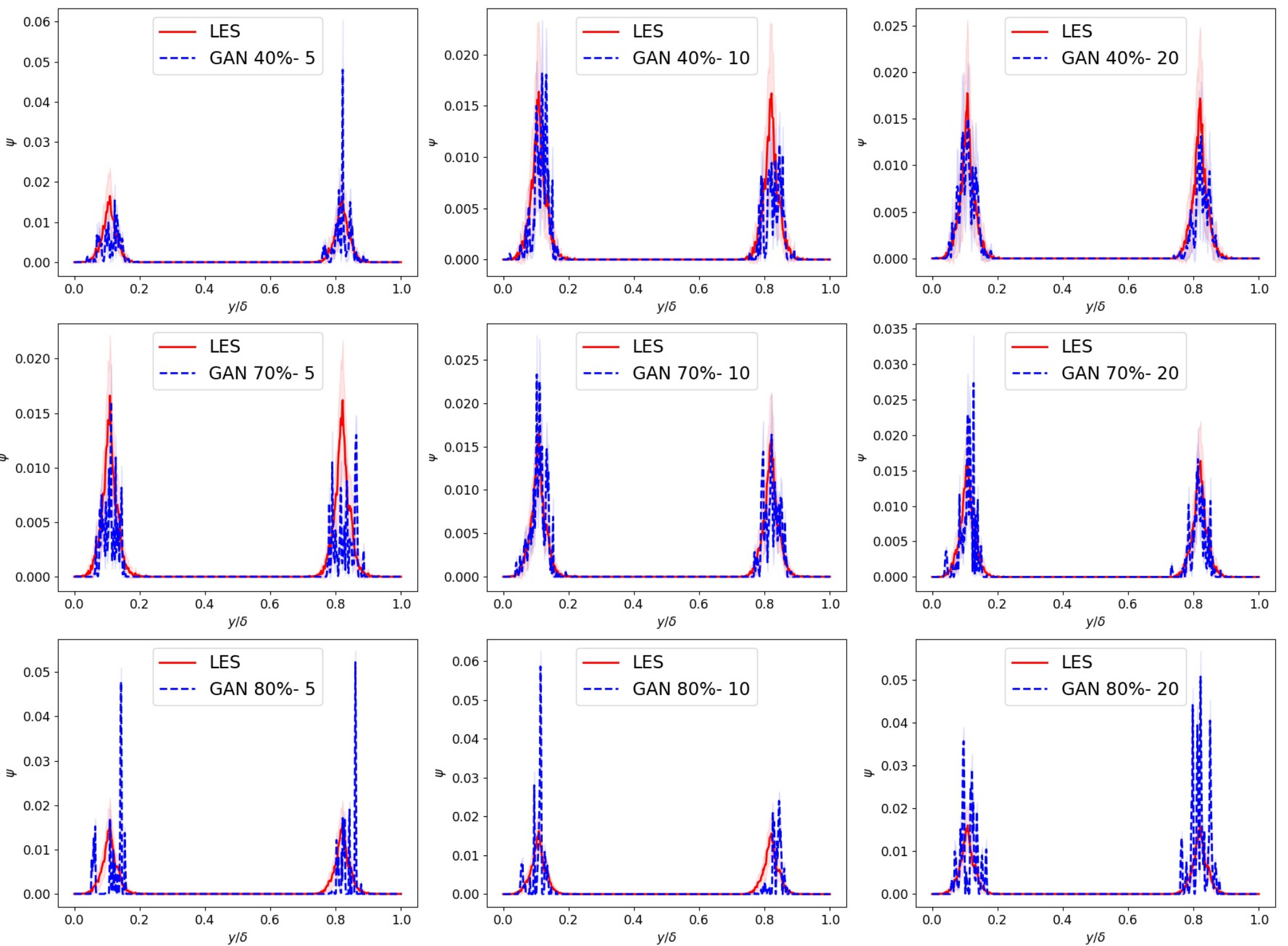}
    \caption{Comparison of the moving average of the velocity field $z$-component for $w(\xi,t)>0$ for LES and GAN synthesized turbulent flow in the rear region of the LPT stator suction side, where the generator $\phi$ is trained on data sets with reduced frame rate at 5, 10 and 20 regular intervals, given a certain percentage of data to be excluded. The red and blue shaded areas indicate the $95\%$ confidence intervals of the respective curve}
    \label{fig:ma-reduction-framerate}
\end{figure}

For the GAN synthesized turbulent flow in the rear region of the LPT we can observe a similar behavior. \Cref{fig:results-reduction-framerate-back} shows that the quality of the GAN synthesized images increases by increasing the number of intervals, especially when the training data is reduced by 60\% and 80\%. Comparing the moving average $\psi$ of the velocity field $z$-component for $w(\xi,t)>0$ shows that the GAN synthesized turbulent flow is in excellent agreement with that of the LES using an interval number of 20 when the training data is reduced by less than 80\% (see \cref{fig:ma-reduction-framerate}). Analogous to the correlation comparison, the curves for $\psi$ of the GAN synthesized turbulence compared to the LES are completely different by a reduction of 80\% of the training data.

\subsection{Computational cost}\label{ssec:Computational-cost-gan}
The GAN training and inference are performed on a GPU of type NVIDIA A100 with 80 GB. For the first setup of the generalization experiments, where we reduce the training data by excluding sequential frames we reach the limit for generating high quality turbulence at a $30\%$ reduction of the training data. Training the \texttt{pix2pixHD} using $75\%$ of the original training data takes $12$ minutes per epoch, resulting in a total of $1.67$ training days in total. For the second setup of our experiments, where we reduce the training data by reducing the frame rate at regular intervals we are able to generate high quality turbulence up to an extraction of $70\%$ of the training data. Training on only $30\%$ of the original training data results in a training time of $6$ minutes per epoch and thus less than a day of training. For comparison, \texttt{pix2pixHD} training on all of the $2,250$ images takes about $15$ minutes per epoch, or more than two days in total. 
The pure inference time for all trained generators is $0.01$ seconds per frame. Thus, the production of $2,250$ frames of the LPT stator under periodic wake impact takes about $22.5$ seconds of inference with the \texttt{pix2pixHD}.

\section{CONCLUSION AND OUTLOOK}
\label{sec:Conclusion_and_outlook}

We demonstrated the capabilities and identified the limits for the generalization of the conditional GAN \texttt{pix2pixHD} considering the test case of flow around an LPT stator under periodic wake impact.
We found that it is possible to reduce the required amount of training data by 70\% by reducing the frame rate at 20 regular intervals. As a result, we do not need to simulate as much data for training, which saves computational time for simulation. We also save computation time by reducing the training time of the GAN by more than 50\% with less data.
In conclusion, we are able to generate high quality turbulence in a complex environment with the conditional GAN, which is excellently matched to the one of the LES visually and in terms of its statistical properties, while significantly reducing the computational time for the training and the generation of the turbulence compared to the LES.

We have shown that conditional GAN has great potential for generalization, but also that the generalization has limits if the change in geometry is too far away from the data used in GAN training. This raises further interesting questions. In particular, it is of interest whether it can be also generalized over the design of the rotor blades. Furthermore, the statistical properties of the GAN synthesized turbulence are in reasonable agreement with the LES turbulence up to a limit, without introducing any physical information into the GAN framework. The next steps are to incorporate physical parameters into the GAN training in the sense of \cite{kim2021} with the aim to further improve the quality of the generated turbulence and to investigate if the limitations regarding the generalizability can be positively influenced by feedback from physics.

\end{document}